%% file: main.tex
\documentclass{vgtc}                      

\onlineid{0}

\vgtccategory{Research}

\vgtcpapertype{please specify}

\title{Urban Computing for Climate and Environmental Justice:\\ Early Perspectives From Two Research Initiatives}

\author{%
  Carolina Veiga\thanks{email: carolvfs@illinois.edu}, Ashish Sharma\thanks{email: sharmaa@illinois.edu}\\\scriptsize University of Illinois \and Daniel de Oliveira\thanks{email: danielcmo@ic.uff.br}, Marcos Lage\thanks{email: mlage@ic.uff.br}\\\scriptsize Universidade Federal Fluminense
  \and Fabio Miranda\thanks{email: fabiom@uic.edu}\\\scriptsize University of Illinois Chicago
}

\abstract{%
The impacts of climate change are intensifying existing vulnerabilities and disparities within urban communities around the globe, as extreme weather events, including floods and heatwaves, are becoming more frequent and severe, disproportionately affecting low-income and underrepresented groups.
Tackling these increasing challenges requires novel approaches that integrate expertise across multiple domains, including computer science, engineering, climate science, and public health.
Urban computing can play a pivotal role in these efforts by integrating data from multiple sources to support decision-making and provide actionable insights into weather patterns, infrastructure weaknesses, and population vulnerabilities.
However, the capacity to leverage technological advancements varies significantly between the Global South and Global North.
In this paper, we present two multiyear, multidisciplinary projects situated in Chicago, USA and Niter\'{o}i, Brazil, highlighting the opportunities and limitations of urban computing in these diverse contexts.
Reflecting on our experiences, we then discuss the essential requirements, as well as existing gaps, for visual analytics tools that facilitate the understanding and mitigation of climate-related risks in urban environments.
}


\renewcommand{\manuscriptnotetxt}{}

\graphicspath{{figs/}{figures/}{pictures/}{images/}{./}} 

\usepackage{tabu}                      
\usepackage{booktabs}                  
\usepackage{lipsum}                    
\usepackage{mwe}                       
\RequirePackage[svgnames]{xcolor}
\usepackage{tikz}

\usepackage{mathptmx}                  

\definecolor{myblue}{HTML}{EEEEFF}
\newcommand*\circled[1]{{\small\tikz[baseline=(char.base)]{\node[shape=circle,fill=myblue,draw,inner sep=1pt] (char) {#1};}}}

\begin{document}

\input{01-introduction}

\input{02-background}

\input{03-niteroi}
\input{04-chicago}
\input{05-discussion}

\bibliographystyle{abbrv-doi-hyperref}
\bibliography{main}

\end{document}

%% file: 01-introduction.tex
\firstsection{Introduction}
\label{sec:intro}
\maketitle


With growing economic and environmental pressures, cities are actively seeking innovative solutions to address current and future challenges.
Over the past decade, urban sensing initiatives and increasingly more capable computing infrastructures have created opportunities for experts from diverse domains, such as engineering, public health, urban planning, and climate science, to tackle pressing problems by leveraging data and analytics.
Such an \emph{urban computing} approach~\cite{ZhengEtAl2014}, in which urban data and computation are leveraged to tackle urban problems, has been instrumental in providing solutions for better disaster management~\cite{AqibEtAl2020}, understanding air and noise pollution patterns~\cite{BelloEtAl2019}, improving mobility and accessibility~\cite{HosseiniEtAl2023}, enhancing city management~\cite{LyuEtAl2024}, and elevating the overall quality of urban life~\cite{MirandaEtAl2019}.

Urban computing, however, presents numerous technological and research challenges spanning the entire data lifecycle: from data acquisition and collection (e.g., how to design low-cost but reliable sensors~\cite{KristoferEtAl2021}, find and organize urban datasets~\cite{CasteloEtAl2021}) to visual analytics (e.g., how to visualize multivariate urban data~\cite{MotaEtAl2023}, automatically extract patterns~\cite{DoraiswamyEtAl2014}).
While significant strides have been made in recent years in democratizing access to some of these technological frameworks, there is still a considerable gap with respect to the accessibility to these advancements across urban areas.
As an inherently multidisciplinary topic, urban computing brings together experts in computing as well as urban problems and, as many topics of such a type, outcomes are heavily influenced by domain differences and priorities, leading to siloed and one-off projects that are rarely translated to other contexts.
Adding to that, social, economic, political, and technological contexts vary from region to region and at various levels: across neighborhoods of the same city~\cite{DaeppEtAl2023}, among cities in the same country~\cite{BarbosaEtAl2014}, and between countries globally~\cite{BiljeckiEtAl2021}.

This paper then proposes a reflection on these challenges, using two multi-year multidisciplinary projects as grounding examples, each leveraging urban computing to tackle different aspects of climate and environmental justice.
The projects are situated in Niterói, Brazil and Chicago, USA -- large cities in different hemispheres that share similar challenges, such as heat waves and flooding, yet exhibit markedly distinct realities.
In Niterói, our team has been working with public officials and climate scientists to leverage heterogeneous urban data to assist the municipal government in analyzing previous rainfalls and their impacts (such as landslides and floods), ultimately enabling data-driven decision-making.
In Chicago, our team has been working with disproportionately impacted communities, including low-income communities and communities of color, as well as with climate scientists in creating tools that highlight and explain environmental and climate injustices.
In common, both projects leverage visualization and visual analytics as a means to facilitate collaboration across various domains and stakeholders.
In this paper, we aim to share our insights from these two projects, reflecting upon the stakeholders' profiles, requirements, urban computing strategies adopted, limitations, and lessons learned in each scenario.
By engaging in this discussion, we hope to provide a pathway to enhance the understanding of how urban computing can contribute to developing healthy and equitable cities.

In Section~\ref{sec:background}, we provide a brief overview of core concepts underlying both projects: urban computing, and climate and environmental justice.
In Section~\ref{sec:examples}, we discuss both projects, as well as their urban computing requirements, particularly focusing on visual analytics.
In Section~\ref{sec:discussion}, we provide a discussion on the technical limitations we faced in both projects.

%% file: 02-background.tex
\section{Background}
\label{sec:background}

In this section, we provide an overview of urban computing, particularly focusing on its connection to visualization and visual analytics.
We then examine climate and environmental justice and their relevance to the two previously mentioned projects.

\begin{figure*}[h!]
    \centering
    \includegraphics[width=1\linewidth]{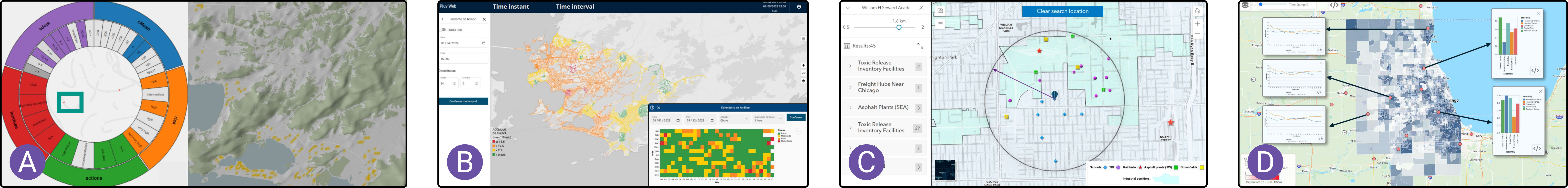}
    \caption{Examples of tools developed as part of the two projects. For Niterói, we have developed RiskVis~\cite{BonadiaEtAl2023} and PluvWeb. For Chicago, we have developed (c) the Proximity-to-Hazard dashboard~\cite{BodaEtAl2023} and (d) the E-Just toolkit~\cite{SharmaEtAl2024}.}
    \label{fig:examples}
\end{figure*}

\subsection{Urban Computing}

Although the concept of \emph{urban computing} originated in works from the 2000s~\cite{KindbergEtAl2007}, Zheng et al. were the first to propose a general urban computing framework~\cite{ZhengEtAl2014}.
In their framework, the components of urban computing encompass the entire data lifecycle.
%
%
Moreover, unlike more targeted systems that are based on a single data type and task, urban data is heterogeneous, spanning, for example, collections of images~\cite{MirandaEtAl2020}, timeseries~\cite{MirandaEtAl2018}, and geo-referenced data~\cite{HuangWang2020}.
Also, tasks can be accomplished by combining different data, techniques, and components within an urban computing framework~\cite{ZhengEtAl2014}.
Next, we will briefly elaborate on these components and revisit them when discussing the projects.

\noindent \circled{C1} \textbf{Collection, generation \& discovery.}
Urban data can be generated through different means, such as sensing~\cite{BelloEtAl2019}, crowdsourcing~\cite{SahaEtAl2019}, or simulations~\cite{YuEtAl2024}.
Given the widespread popularity of open data portals, data discovery also plays an important role in urban computing~\cite{CatlettEtAl2014}.

\noindent \circled{C2} \textbf{Curation \& transformation.}
With the growth in the availability of urban data, organizing and maintaining datasets is also an important component of urban computing.
This also includes cleaning~\cite{ZhaoEtAl2016} and creating appropriate indices~\cite{DoraiswamyEtAl2018} to organize and facilitate search by different users and stakeholders.

\noindent \circled{C3} \textbf{Management.}
The complexity and size of urban data require the use of non-trivial approaches to enable scalability and support interactive exploratory analyses~\cite{DoraiswamyEtAl2018}.

\noindent \circled{C4} \textbf{Analysis \& modeling.}
Due to the scale of the data, both the size of individual datasets and the number of datasets, manual exploration is often not practical.
Recent approaches in urban computing use machine learning~\cite{XieEtAl2020} and computational topology~\cite{MirandaEtAl2017} to identify interesting features and events in one dataset~\cite{DoraiswamyEtAl2014}, or relationships between datasets~\cite{ChirigatiEtAl2016}.

\noindent \circled{C5} \textbf{Visualization.}
Lastly, visualization plays a crucial role in making sense of complex urban data. Effective techniques enable researchers and domain experts to understand trends and anomalies within the data, considering 2D~\cite{ZhengEtAl2016,DengEtAl2023} and 3D data~\cite{MirandaEtAl2024}.

\subsection{Climate and Environmental Justice}

The topics of climate and environmental justice are also integral to both projects.
The social movements that emerged in the 1960s, along with the increased frequency of natural disasters, has underscored the need for climate and environmental justice in urban areas.
These concepts consider that many of those most severely impacted are vulnerable minorities who have had minimal influence over the processes and activities that contribute to their adverse circumstances, and on the related decision-making. 
In essence, climate change and environmental problems are closely linked to social, ethnic, political, and economic issues. 
%
%
Our primary objective is to present two research initiatives tackling climate and environmental justice through a participatory approach that involves interaction and collaboration among diverse stakeholders. These projects are grounded in the urban computing concepts discussed earlier.
Although we discuss the projects in broad terms, we cite research articles that provide more detailed descriptions of some of the outcomes.

%% file: 03-niteroi.tex
\section{Examples of Urban Computing Projects}
\label{sec:examples}
In this section, we provide two examples of urban computing projects, each with different requirements and limitations. 

\vspace{-0.2cm}
\subsection{Niterói: Flooding \& Landslides Disaster Management}
\label{sec:niteroi}
Niterói, a medium-sized city in Brazil, is situated in the mountainous regions of Rio de Janeiro. The city frequently endures climate-related disasters, such as landslides and floods~\cite{SmythRoyle2000}.
In 2019, the city launched a call for research projects that proposed to develop solutions focused on sustainability.
The project described next was one of the funded projects and aimed at developing solutions to aid the city's Civil Defense Department in analyzing rainfall data, as well as related events, such as landslides and floods.
Some of the components of the project build on top of our previous efforts leveraging weather simulations for disaster management~\cite{DeSouzaEtAl2022, DeSouzaEtAl2024, BonadiaEtAl2023}.

Throughout the three-year project, we held several meetings with professionals from the city's department. In this process, we were able to design and implement our own urban computing framework, leveraging both off-the-shelf and custom-built technologies for each one of the \circled{C1}-\circled{C5} components.
During the initial meetings, we began collecting and curating a set of heterogeneous datasets. For \circled{C1}, we then collected observed data (e.g., accumulated rain volumes, landslide and flood occurrences, weather warnings and alert messages, traffic camera videos), as well as sociodemographic data.
For \circled{C2}, the project required the consolidation of the spatial and temporal data. To achieve this, data transformations included the extraction of frames from the video data, as well as spatiotemporal computations, such as computing accumulated rainfall volumes.
For \circled{C3}, we made use of spatial indices to support interactive data exploration, including the comparison of spatial and temporal patterns. Additionally, we dedicated significant effort to ensure the accurate integration of all relevant data.
In specific, we leveraged our previous work~\cite{DeSouzaEtAl2022, DeSouzaEtAl2024, BonadiaEtAl2023} that proposed a data integration framework that takes data from multiple sources with different mathematical descriptions, dimensions, and resolutions and creates a data-enriched triangle mesh that can be used for both visualization and numerical simulation applications.

In our early communications with city officials, it was made clear to us that the dataset with flood occurrences was not as comprehensive as desired, with many spatial and temporal gaps. As part of \circled{C4}, we then decided to leverage the video data to train a computer vision model to detect flood hotspots.
Another analytical component was the ability to compute response route options for emergency vehicles, taking into account traffic and flood data.
For \circled{C5}, data and analyses were made available through a visual interface that used well-known visual designs, such as heatmaps and line charts.







%% file: 04-chicago.tex
\vspace{-0.2cm}
\subsection{Chicago: Weather \& Climate Analysis}
\label{sec:chicago}

Over the past few years, we have partnered with a number of coalition groups in Chicago to develop dashboard-like visualization interfaces to investigate climate and environmental justice issues~\cite{SharmaEtAl2024}.
These communities are largely Latino and African American and have per capita income levels that are lower than Chicago's average.
Due to the industrial zone land use classification and proximity to major highways, the number of facilities in the surrounding industrial corridors is increasing.

As part of these ongoing projects, we employed a participatory approach~\cite{BodaEtAl2023}, in which many of the problems and data included in our outcomes were identified and suggested by community members.
Similar to the Niterói project, \circled{C1} relied on crowdsourced, observed and sociodemographic data.
For crowdsourced data, we relied on OpenStreetMap for building and street network information.
For observed data, for example, we included data from smart city initiatives in Chicago (e.g., Array of Things) as well as ECOSTRESS and MODIS space sensors describing temperature and air pollution~\cite{SayedEtAl2024}. For sociodemographic data, we included average income, race, and ethnicity information made available by the US Census.
For simulated data, our tools incorporated weather simulation data from WRF-Chem.
We also included toxic release inventory facilities made available by the US EPA.

For \circled{C2}, we transformed and filtered space sensors data from GeoTIFF to NetCDF format, and filtered WRF-Chem data to include only variables of interest. Observed data were aggregated hourly, and sociodemographic data was structured according to each spatial unit. Additionally, we aggregated data by region, time window, and variable to facilitate easier comparisons across datasets.
For \circled{C3}, processed files are stored as JSON files and made available through a server to the web client.

For \circled{C4}, we aimed to support different tasks, such as computing the correlation between variables across spatial units, comparing observed and simulated values over time, and calculating forecast verification metrics.
Lastly, for \circled{C5}, given the need to support users across domains and visual literacy levels, we designed a grammar-based framework that enabled us to quickly iterate over different visualization designs. We can customize the interface to align with the data and the user profile. This process involves selecting and organizing components such as maps and charts. Each map layer is created based on the data and specified grammar, representing grids, regions, spatial points, and 3D buildings. Additionally, the grammar defines the interactions for each component, allowing for flexible and tailored visual analyses of specific areas or elements according to the data and user requirements.
Using the custom grammar, we can create individual and juxtaposed views for spatial comparisons and statistical analyses. These analyses focused on specific regions or time windows, usually when the overview map indicated relevant patterns.





%% file: 05-discussion.tex
\section{Discussion \& Conclusion}
\label{sec:discussion}

Through our experience in developing visual analytics tools that integrate urban computing components and leverage extensive data from two different locations, we were able to better understand limitations, not only from a research perspective but also a practical one.
First, we share our perspectives considering each urban computing component. Then, we outline potential future work.

In \circled{C1}, collecting data from multiple sources demands significant effort, especially since many datasets, such as crowdsourced or sensor data, are continuously updated. This requires either manually downloading up-to-date data or implementing streaming capabilities and API connections with appropriate data sources. In our experience, there is a considerable barrier to leveraging sensing data, in particular. In Niterói, even though the city had access to a number of cameras, creating appropriate data pipelines to extract and process data for the analytical process was a major hurdle. We faced similar challenges when leveraging simulation data, which requires a set of complex steps to format data for visualization.

In \circled{C2}, we have observed two challenges. First, data size can make spatial joins prohibitively expensive, particularly when aggregating over street networks. Second, changes in requirements, either arising from new interviews with experts or data limitations, require modifying already existing data workflows. Given the complexity of some of these workflows, spread over potentially multiple computational notebooks, adjustments can be time-consuming and prone to errors.
In \circled{C3}, despite significant advancements in new indices and techniques for spatial data in recent years, most of these innovations remain confined to tabular data. Considering the complexity of urban data and tasks, there is still a gap in addressing advanced data types, such as images, networks, and geometries.

In \circled{C4}, given urban computing's societal implications, ensuring the transparency, interpretability, and accuracy of data-driven methodologies is a fundamental concern. This has also been highlighted in previous works, particularly with respect to environmental justice~\cite{RivkahEtAl2022}. In the Niterói project, where we had to train a new flood detection computer vision model, there were concerns with the downstream reliability of the approach, particularly considering different weather conditions.

In \circled{C5}, challenges centered on maintaining a connection between data and visualizations. Given data and workflow complexities, early commitments to certain visualization designs were common. Moreover, the creation of visual analytics tools for urban computing requires the integration of multiple components. Changes in requirements often entail substantial redevelopment efforts.

\noindent \textbf{Takeaways \& research opportunities.}
As outlined in previous surveys~\cite{XieEtAl2020, DengEtAl2023, FerreiraEtAl2024, MirandaEtAl2024}, urban computing presents numerous research opportunities for visualization.
Our experience with these two projects (as well as previous ones) reveals that a pressing challenge remains in moving beyond siloed efforts and ensuring interoperability of outcomes.
As of now, properly setting up all components of an urban computing data pipeline for a visual analytics tool requires considerable effort, given data and task complexities.
This leads to several consequences, including the creation of monolithic prototypes and poor reusability of outcomes. Most importantly, there should be a greater focus on moving beyond one-off collaborations and ensuring that research outcomes from one project can be \emph{easily} leveraged by others.
While there have been initial efforts in this direction (such as Urbanity~\cite{YapEtAl2023} or our own the Urban Toolkit~\cite{MoreiraEtAl2024}), the \emph{status quo} is one where visual analytics tools and applications are still largely being built from scratch, with little to no concerns for reusability or interoperability.
%
We argue for stronger efforts to enhance interoperability, allowing outcomes to be easily utilized across the urban computing pipeline and across projects.
Visual analytics researchers are uniquely qualified for such endeavors, as the tools and applications developed by the community often require a comprehensive understanding of the entire pipeline.

On top of that, we also see opportunities to more inclusively account for experts' existing workflows. With the growing popularity of data science libraries and toolkits, urban experts are increasingly adopting them into their practice.
In turn, they are becoming more aware (and proficient) of  approaches that require computer science expertise.
For example, in climate science, computational notebooks are becoming increasingly used for data exploration, despite their well-known limitations~\cite{PimentelEtAl2019}.
In our projects, we invested significant effort in understanding these notebooks and transforming them into components better suited to our visualization objectives -- ultimately replacing them with bespoke tools.
Rather than replacing these workflows, we see an opportunity for new approaches that can bridge the gap between diverse data sources, analytical approaches, and interactive environments.
Such an approach would preserve the familiarity of computational notebooks and also leverage visualization strengths to create more robust and scalable tools.

Lastly, building on the previous points, we see a significant opportunity to facilitate collaborative efforts across domains. In particular, fostering collaborative approaches that leverage interoperability and new interactive environments could lower the barrier to the iterative design of urban computing dataflows. In such a scenario, changes made by experts in any component of a dataflow could be easily propagated to subsequent stages of the pipeline, avoiding the need for re-design and re-implementation efforts.
%

\vspace{-0.2cm}
\section*{Acknowledgments}
\vspace{-0.2cm}
This study was supported by NASA (\#80NSSC22K1683), NSF (\#2320261, \#2330565, \#2411223, \#2139316, \#2230772), IDOT (TS-22-340), CNPq (316963/2021-6, 311425/2023-2), and FAPERJ (E-26/202.915/2019, E-26/211.134/2019).
\vspace{-0.3cm}

%% file: main.bbl
\begin{thebibliography}{10}

\bibitem{AqibEtAl2020}
M.~Aqib, R.~Mehmood, A.~Alzahrani, and I.~Katib.
\newblock {\em A Smart Disaster Management System for Future Cities Using Deep Learning, GPUs, and In-Memory Computing}, pp. 159--184.
\newblock Springer International Publishing, 2020.

\bibitem{BarbosaEtAl2014}
L.~Barbosa, K.~Pham, C.~Silva, M.~R. Vieira, and J.~Freire.
\newblock Structured open urban data: Understanding the landscape.
\newblock {\em Big Data}, 2(3):144--154, 2014.
\newblock PMID: 25276498.

\bibitem{BelloEtAl2019}
J.~P. Bello, C.~Silva, O.~Nov, R.~L. Dubois, A.~Arora, J.~Salamon, C.~Mydlarz, and H.~Doraiswamy.
\newblock {SONYC}: a system for monitoring, analyzing, and mitigating urban noise pollution.
\newblock {\em Commun. ACM}, 62(2):68–77, 2019.

\bibitem{BiljeckiEtAl2021}
F.~Biljecki, L.~Z.~X. Chew, N.~Milojevic-Dupont, and F.~Creutzig.
\newblock Open government geospatial data on buildings for planning sustainable and resilient cities.
\newblock {\em Arxiv}, 2021.

\bibitem{BodaEtAl2023}
P.~A. Boda, F.~Fusi, F.~Miranda, et~al.
\newblock Environmental justice through community-policy participatory partnerships.
\newblock {\em Journal of Environmental Protection}, 14:616--636, 2023.

\bibitem{BonadiaEtAl2023}
S.~Bonadia, R.~Gama, D.~de~Oliveira, F.~Miranda, and M.~Lage.
\newblock Visual analytics using heterogeneous urban data.
\newblock In {\em 2023 36th SIBGRAPI Conference on Graphics, Patterns and Images}, pp. 1--6, 2023.

\bibitem{CasteloEtAl2021}
S.~Castelo, R.~Rampin, A.~Santos, A.~Bessa, F.~Chirigati, and J.~Freire.
\newblock Auctus: a dataset search engine for data discovery and augmentation.
\newblock {\em Proc. VLDB Endow.}, 14(12):2791–2794, 2021.

\bibitem{CatlettEtAl2014}
C.~Catlett, T.~Malik, B.~Goldstein, et~al.
\newblock Plenario: An open data discovery and exploration platform for urban science.
\newblock {\em IEEE Data Eng. Bull.}, 37(4):27--42, 2014.

\bibitem{KristoferEtAl2021}
K.~Chan, D.~N. Schillereff, A.~C. Baas, M.~A. Chadwick, B.~Main, M.~Mulligan, F.~T. O’Shea, R.~Pearce, T.~E. Smith, A.~van Soesbergen, E.~Tebbs, and J.~Thompson.
\newblock Low-cost electronic sensors for environmental research: Pitfalls and opportunities.
\newblock {\em Progress in Physical Geography: Earth and Environment}, 45(3):305--338, 2021.

\bibitem{ChirigatiEtAl2016}
F.~Chirigati, H.~Doraiswamy, T.~Damoulas, and J.~Freire.
\newblock Data polygamy: The many-many relationships among urban spatio-temporal data sets.
\newblock In {\em Proc. of the 2016 International Conference on Management of Data}, SIGMOD '16, p. 1011–1025, 2016.

\bibitem{DaeppEtAl2023}
M.~I.~G. Daepp, A.~Cabral, T.~M. Werner, R.~Mansour, C.~Catlett, A.~Roseway, C.~Needham, N.~Udeagbala, and S.~Counts.
\newblock The “three-legged stool": Designing for equitable city, community, and research partnerships in urban environmental sensing.
\newblock In {\em Proc. of the 2023 CHI Conference on Human Factors in Computing Systems}, CHI '23, 2023.

\bibitem{DeSouzaEtAl2024}
C.~V.~F. de~Souza, S.~M. Bonnet, D.~de~Oliveira, M.~Cataldi, F.~Miranda, and M.~Lage.
\newblock Prowis: A visual approach for building, managing, and analyzing weather simulation ensembles at runtime.
\newblock {\em IEEE Trans. Vis. Comput. Graph.}, 30(1):738--747, 2024.

\bibitem{DeSouzaEtAl2022}
C.~V.~F. {de Souza}, P.~da~Cunha~{Luz Barcellos}, L.~Crissaff, M.~Cataldi, F.~Miranda, and M.~Lage.
\newblock Visualizing simulation ensembles of extreme weather events.
\newblock {\em Computers \& Graphics}, 104:162--172, 2022.

\bibitem{DengEtAl2023}
Z.~Deng, D.~Weng, S.~Liu, Y.~Tian, M.~Xu, and Y.~Wu.
\newblock A survey of urban visual analytics: Advances and future directions.
\newblock {\em Comp. Visual Media}, 9:3--39, 2023.

\bibitem{DoraiswamyEtAl2014}
H.~Doraiswamy, N.~Ferreira, T.~Damoulas, J.~Freire, and C.~T. Silva.
\newblock Using topological analysis to support event-guided exploration in urban data.
\newblock {\em IEEE Trans. Vis. Comput. Graph.}, 20(12):2634--2643, 2014.

\bibitem{DoraiswamyEtAl2018}
H.~Doraiswamy, E.~T. Zacharatou, F.~Miranda, M.~Lage, A.~Ailamaki, C.~T. Silva, and J.~Freire.
\newblock Interactive visual exploration of spatio-temporal urban data sets using urbane.
\newblock In {\em Proc. of the 2018 International Conference on Management of Data}, SIGMOD '18, p. 1693–1696, 2018.

\bibitem{FerreiraEtAl2024}
L.~Ferreira, G.~Moreira, M.~Hosseini, M.~Lage, N.~Ferreira, and F.~Miranda.
\newblock Assessing the landscape of toolkits, frameworks, and authoring tools for urban visual analytics systems.
\newblock {\em Computers \& Graphics}, p. 104013, 2024.

\bibitem{RivkahEtAl2022}
R.~Gardner-Frolick, D.~Boyd, and A.~Giang.
\newblock Selecting data analytic and modeling methods to support air pollution and environmental justice investigations: A critical review and guidance framework.
\newblock {\em Environmental Science \& Technology}, 56(5):2843--2860, 2022.

\bibitem{HosseiniEtAl2023}
M.~Hosseini, A.~Sevtsuk, F.~Miranda, R.~M. Cesar, and C.~T. Silva.
\newblock Mapping the walk: A scalable computer vision approach for generating sidewalk network datasets from aerial imagery.
\newblock {\em Computers, Environment and Urban Systems}, 101:101950, 2023.

\bibitem{HuangWang2020}
B.~Huang and J.~Wang.
\newblock Big spatial data for urban and environmental sustainability.
\newblock {\em Geo-spatial Information Science}, 23(2):125--140, 2020.

\bibitem{KindbergEtAl2007}
T.~Kindberg, M.~Chalmers, and E.~Paulos.
\newblock Guest editors' introduction: Urban computing.
\newblock {\em IEEE Pervasive Computing}, 6(3):18--20, 2007.

\bibitem{LyuEtAl2024}
Y.~Lyu, H.~Lu, M.~K. Lee, G.~Schmitt, and B.~Y. Lim.
\newblock {IF-City}: Intelligible fair city planning to measure, explain and mitigate inequality.
\newblock {\em IEEE Trans. Vis. Comput. Graph.}, 30(7):3749--3766, 2024.

\bibitem{MirandaEtAl2019}
F.~Miranda, H.~Doraiswamy, M.~Lage, L.~Wilson, M.~Hsieh, and C.~T. Silva.
\newblock {Shadow Accrual Maps}: Efficient accumulation of city-scale shadows over time.
\newblock {\em IEEE Trans. Vis. Comput. Graph.}, 25(3):1559--1574, 2019.

\bibitem{MirandaEtAl2017}
F.~Miranda, H.~Doraiswamy, M.~Lage, K.~Zhao, B.~Gonçalves, L.~Wilson, M.~Hsieh, and C.~T. Silva.
\newblock {Urban Pulse}: Capturing the rhythm of cities.
\newblock {\em IEEE Trans. Vis. Comput. Graph.}, 23(1):791--800, 2017.

\bibitem{MirandaEtAl2020}
F.~Miranda, M.~Hosseini, M.~Lage, H.~Doraiswamy, G.~Dove, and C.~T. Silva.
\newblock {Urban Mosaic}: Visual exploration of streetscapes using large-scale image data.
\newblock In {\em Proc. of the 2020 CHI Conference on Human Factors in Computing Systems}, CHI '20, p. 1–15, 2020.

\bibitem{MirandaEtAl2018}
F.~Miranda, M.~Lage, H.~Doraiswamy, C.~Mydlarz, J.~Salamon, Y.~Lockerman, J.~Freire, and C.~T. Silva.
\newblock {Time Lattice}: A data structure for the interactive visual analysis of large time series.
\newblock {\em Computer Graphics Forum}, 37(3):23--35, 2018.

\bibitem{MirandaEtAl2024}
F.~Miranda, T.~Ortner, G.~Moreira, M.~Hosseini, M.~Vuckovic, F.~Biljecki, C.~T. Silva, M.~Lage, and N.~Ferreira.
\newblock The state of the art in visual analytics for {3D} urban data.
\newblock {\em Computer Graphics Forum}, 43(3):e15112, 2024.

\bibitem{MoreiraEtAl2024}
G.~Moreira, M.~Hosseini, M.~N.~A. Nipu, M.~Lage, N.~Ferreira, and F.~Miranda.
\newblock {The Urban Toolkit}: A grammar-based framework for urban visual analytics.
\newblock {\em IEEE Trans. Vis. Comput. Graph.}, 30(1):1402--1412, 2024.

\bibitem{MotaEtAl2023}
R.~Mota, N.~Ferreira, J.~D. Silva, M.~Horga, M.~Lage, L.~Ceferino, U.~Alim, E.~Sharlin, and F.~Miranda.
\newblock A comparison of spatiotemporal visualizations for {3D} urban analytics.
\newblock {\em IEEE Trans. Vis. Comput. Graph.}, 29(1):1277--1287, 2023.

\bibitem{PimentelEtAl2019}
J.~F. Pimentel, L.~Murta, V.~Braganholo, and J.~Freire.
\newblock A large-scale study about quality and reproducibility of {Jupyter Notebooks}.
\newblock In {\em {IEEE}/{ACM} {International} {Conference} on {Mining} {Software} {Repositories}}, pp. 507--517, 2019.

\bibitem{SahaEtAl2019}
M.~Saha, M.~Saugstad, H.~T. Maddali, et~al.
\newblock {Project Sidewalk}: A web-based crowdsourcing tool for collecting sidewalk accessibility data at scale.
\newblock In {\em Proc. of the 2019 CHI Conference on Human Factors in Computing Systems}, CHI '19, p. 1–14, 2019.

\bibitem{SayedEtAl2024}
T.~K. Sayyed, U.~Ovienmhada, M.~Kashani, et~al.
\newblock Satellite data for environmental justice: a scoping review of the literature in the united states.
\newblock {\em Environmental Research Letters}, 19(033001), 2024.

\bibitem{SharmaEtAl2024}
A.~Sharma, C.~Veiga, P.~Li, et~al.
\newblock {e-JUST}-environmental justice using urban scalable toolkit.
\newblock In {\em 104th AMS Annual Meeting}. AMS, 2024.

\bibitem{SmythRoyle2000}
C.~G. Smyth and S.~A. Royle.
\newblock Urban landslide hazards: incidence and causative factors in {Niterói, Rio de Janeiro State, Brazil}.
\newblock {\em Applied Geography}, 20(2):95--118, 2000.

\bibitem{XieEtAl2020}
P.~Xie, T.~Li, J.~Liu, S.~Du, X.~Yang, and J.~Zhang.
\newblock Urban flow prediction from spatiotemporal data using machine learning: A survey.
\newblock {\em Information Fusion}, 59:1--12, 2020.

\bibitem{YapEtAl2023}
W.~Yap, R.~Stouffs, and F.~Biljecki.
\newblock Urbanity: automated modelling and analysis of multidimensional networks in cities.
\newblock {\em npj Urban Sustain}, 3(45), 2023.

\bibitem{YuEtAl2024}
Y.~Yu, P.~Li, D.~Huang, and A.~Sharma.
\newblock Street-level temperature estimation using graph neural networks: Performance, feature embedding and interpretability.
\newblock {\em Urban Climate}, 56:102003, 2024.

\bibitem{ZhaoEtAl2016}
K.~Zhao, S.~Tarkoma, S.~Liu, and H.~Vo.
\newblock Urban human mobility data mining: An overview.
\newblock In {\em 2016 IEEE International Conference on Big Data (Big Data)}, pp. 1911--1920, 2016.

\bibitem{ZhengEtAl2014}
Y.~Zheng, L.~Capra, O.~Wolfson, and H.~Yang.
\newblock Urban computing: Concepts, methodologies, and applications.
\newblock {\em ACM Trans. on Intelligent Systems and Technology}, 5(3), 2014.

\bibitem{ZhengEtAl2016}
Y.~Zheng, W.~Wu, Y.~Chen, H.~Qu, and L.~Ni.
\newblock Visual analytics in urban computing: An overview.
\newblock {\em IEEE Trans. on Big Data}, 2(3):276--296, 2016.

\end{thebibliography}
